# Multipath Signal-Selective Metasurface: Passive Time-Varying Interlocking Mechanism to Vary Spatial Impedance for Signals with the Same Frequency

Kaito Tachi,[1] Kota Suzuki,[1] Kairi Takimoto,[1] Shunsuke Saruwatari,[2] Kiichi Niitsu,[3] Peter Njogu,[1] and Hiroki Wakatsuchi[1]

[1]*Department of Engineering, Nagoya Institute of Technology, Nagoya, Aichi, 466-8555 Japan*
[2]*Graduate School of Information Science and Technology, Osaka University, Suita, Osaka, 565-0871 Japan*
[3]*Faculty of Engineering, Kyoto University, Kyoto, 606-8501 Japan*

Electromagnetic (EM) multipath interference is difficult to address with passive approaches due to two physical restrictions—the shared frequency of the initial and interfering signals and their variable incident angles. Thus, to address multipath interference, the spatial impedance must be adjusted in response to the incident angles of multiple signals with the same frequency, which is impossible with classic linear time-invariant (LTI) systems. We present a design concept for metasurface-based spatial filters to overcome LTI behavior and suppress multipath interference signals using a time-varying interlocking mechanism without any active biasing systems. The proposed devices are coupled to the first incoming wave to adjust the spatial impedance and suppress delayed waves in the time domain, which is validated numerically and experimentally. This study opens a new avenue for passive yet time-varying selective EM metasystems, enabling the adjustment of spatially complicated EM waves and fields even at the same frequency.

In multipath propagation, radio signals reach receiving antennas through multiple routes [1–3]. As the scattering environment evolves, the amplitude of the received signal varies over time. This phenomenon is called fading [4] and occurs when signals reach the receiving antenna from multiple paths, as the signal waves are not necessarily in phase [5]. This multipath interference leads to reliability issues in applications such as terrestrial television broadcasting, resulting in 'ghosting' phenomena, such as duplicate images in television broadcasting [6] and fading in wireless communication [7].

To address the multipath problem, two physical limitations need to be overcome. First, most materials scatter oscillating electromagnetic (EM) waves based on their frequency components [8–10]. Such materials have linear time-invariant (LTI) responses [11], which ensure a constant scattering profile for a given frequency regardless of the arrival time of the signal. This LTI behavior needs to be addressed in multipath fading scenarios to allow transmission of only the first incoming wave and prevent the transmission of any delayed waves. Importantly, the incident angles of the first wave and other waves are not fixed in most cases and need to be detected separately. Second, the interfering signal has the same frequency as the first transmitted signal. The two signals need to be differentiated, but most classic filtering techniques are based on the frequency and thus not applicable in such scenarios [12]. Therefore, to address the multipath issue, the spatial impedance must be adjusted in response to the incident angle even for signals with the same frequency.

To date, metasurfaces (MSs) [13–15] have been extensively explored to adjust the phase and amplitude of transmitted fields [16–22]. In particular, by using nonlinear components, enhanced tunable responses can be achieved with MSs [23, 24]. By harnessing this advanced capability of MSs with nonlinear components, spatially varying impedance can be achieved. In this work, we propose an MS-based filter design that overcomes the LTI limitations of signals with the same frequency to develop a passive yet autonomous multipath signal suppression paradigm with no postprocessing requirements. The signal propagation can be spatially controlled based on its arrival sequence by introducing an interlocking mechanism, which is achieved via nonlinear electronic components with time-varying responses incorporated within the MS.

The proposed MS enables the suppression of EM interference on the receiving side, similar to spatial filtering of energy from a multipath beam, as illustrated in Fig. 1. Fig. 1(a) shows the conventional multipath scenario without (left) and with the MS filter (right). Fig. 1(b) illustrates the ideal received signal with the proposed filter. Note that the multipath environment in Fig. 1(b) is simplified compared to that in Fig. 1(a) by using two transmitters generating the same signals with and without a delay. In this scenario, the MS filter is designed to permit the transmission of the first wave while preventing that of the second incoming wave. Fig. 2(a) illustrates the conceptual transmission-line equivalent circuit model of the proposed spatial multipath filter with two MSs (MS1 and MS2). The transmitters are represented by AC sources, input impedances $Z_0$, and switches, while the receiver is represented by the shunt impedance $Z_1$. The MSs are modeled as variable shunt impedances $Z_{ms1}$ and $Z_{ms2}$ between the transmission lines representing the wave impedance of free space. As shown in Fig. 2(b), incoming signals can be controlled if the states of $Z_{ms1}$ and $Z_{ms2}$ are switched between an open circuit and a short circuit. In this study, these impedances were specifically determined based on the unit cells of a waveform-selective MS, as briefly reviewed in Supplementary Note A [25–27]. As shown in Fig. 2(c), our unit cell consisted of a 15 mm × 7 mm slit with a gold flash conductor on a 1.27 mm thick Rogers 3010 substrate with a relative permittivity of 3. A capacitor $C_{add}$ was connected across the slit, as depicted in Fig. 2(d), to control the resonance frequency and maintain the subwavelength periodicity. In addition, a set of four diodes (HSMS286x series, Avago) was connected with capacitor $C_1$ and a metal-oxide-semiconductor field-effect transistor (MOSFET) (Toshiba, 2KS1062).

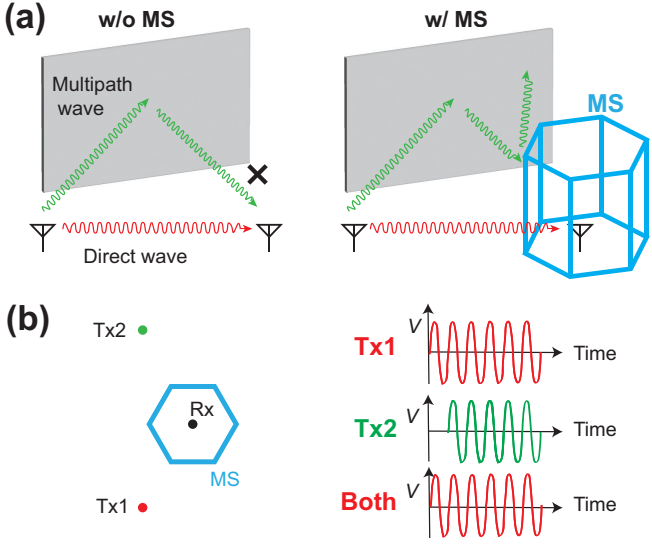

FIG. 1. Conceptual image of the proposed MS-based multipath filter for adjusting spatial wave propagation for signals with the same frequency. (a) Conventional multipath signal environment without (left) and with (right) the proposed MS filter. (b) Simplified equivalent multipath environment (left) and received signals using only Tx1, only Tx2, and both (top right to bottom right). The proposed filter accepts only the Tx1 signal while eliminating the time-delayed multipath signal from Tx2. However, only the Tx2 signal is received if this signal arrives before the Tx1 signal (not shown).

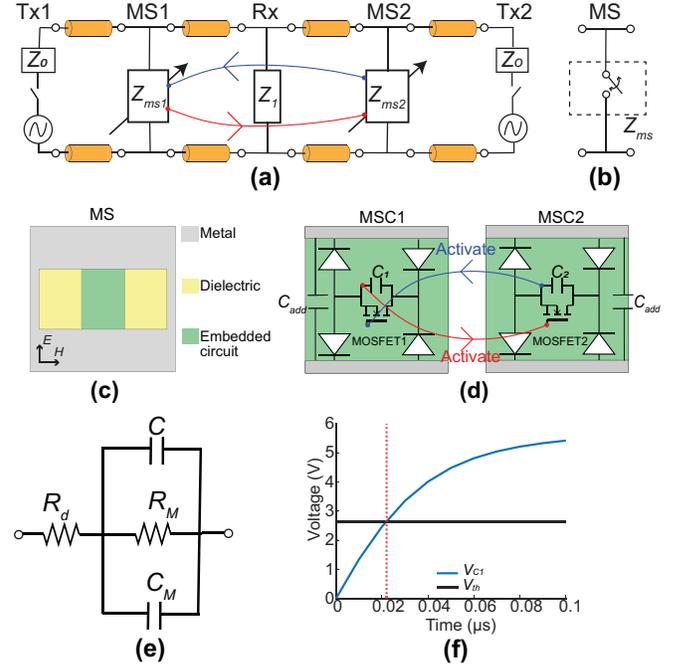

FIG. 2. Operational mechanism of the proposed MS-based filter. (a) The equivalent transmission-line model. The MSs are represented by variable impedances $Z_{ms}$s (i.e., $Z_{ms1}$ and $Z_{ms2}$). (b) Simplified ideal circuit operation of the MSs. (c) The specific MS unit cell design and (d) the circuit components loaded and interlinked to another cell. (e) Simplified equivalent circuit representing the time-domain response and (f) capacitor potential $V_{c1}$ changing over time with the threshold voltage $V_{th}$ of the MOSFET. When $V_{c1}$ is larger than $V_{th}$, MOSFET2 in MSC2 short circuits, preventing the transmission of incoming (delayed multipath) signals. The response time can be reduced by decreasing $C$ (see Supplementary Note A).

In contrast to conventional waveform-selective MSs [25, 26], our MS unit cell includes an MOSFET instead of a resistor. Therefore, the transmission characteristics could be adjusted according to the drain-source resistance $R_M$ of the MOSFET, which varied with the gate-source voltage as follows. When the gate-source voltage $V_g$ was low, $R_M$ became so large that the MS resonance was maintained to strongly transmit incoming signals. However, when $V_g$ was large, $R_M$ became small enough to prevent both the MS resonance and the transmission of incoming waves. According to the proposed spatial filtering approach, two MS circuits (MSC1 and MSC2) were symmetrically connected, as shown in Fig. 2(d). Thus, the potential $V_{C1}$ of capacitor $C_1$ was applied to the gate of MOSFET2 in MSC2. Note that this potential could be obtained when a signal entered from the unit cell of MS1.

Depending on $V_{C1}$, i.e., the presence or absence of a signal entering MS1, the effective resistance $R_{M2}$ between the drain and the source of MOSFET2 varied. With this symmetrical circuit design, the voltage $V_{C2}$ of capacitor $C_2$ could be used to control MOSFET1 in MS1 and its transmittance when a signal enters from MS2. Therefore, this circuit configuration could separately control the transmitting capabilities of the two MS unit cells to suppress only time-delayed signals.

In our filter design, the MS response time $t_{total}$ depends on three factors, which can be summed as $t_{total} = t_1 + t_2 + t_3$. $t_1$, $t_2$, and $t_3$ are needed to generate sufficiently large $V_{C1}$ (or $V_{C2}$), transmit the bias voltage to the neighboring cell, and change the state of the MS unit cell, respectively. In particular, $t_1$ can be estimated according to the simplified equivalent circuit shown in Fig. 2(e). This circuit effectively represents the time-domain response of the circuit with the MS except for $C_{add}$, which is related to only the frequency-domain response [28]. Assuming that a DC voltage source $V_{DC}$ is applied to the equivalent circuit to approximate the rectification process of the MS, $V_{C1}$ can be calculated according to [28]

$$V_{C1}(t) = \frac{R_M V_{DC}(1 - e^{-t/\tau})}{R_M + R_d}, \quad (1)$$

where $t$ and $R_d$ denote time and the resistive component of two diodes at the turn-on voltage (680 Ω in this study), respectively. In addition, $\tau$ is a time constant that characterizes the time-domain response and is obtained as

$$\tau = \frac{C_{all} R_M R_d}{R_M + R_d}, \quad (2)$$

where $C_{all} = C + C_M$ and $C_M$ is the parasitic capacitance of MOSFET1. Note that since $R_M \gg R_d$, $\tau \sim C_{all} R_d$ and $V_{C1}(t) \sim V_{DC}(1 - e^{-t/\tau})$. Thus, when $V_{C1}$ reaches the threshold voltage $V_{th}$ of the MOSFET, i.e., $V_{C1} = V_{th}$, $t_1$ is obtained by

$$t_1 \sim \ln\left(1 - \frac{V_{th}}{V_{DC}}\right)^{-\tau}. \quad (3)$$

Fig. 2(f) shows an example using $C_1 = 100$ pF and $V_{th}$ = 2.65 V. Clearly, Eq. (3) indicates that $t_1$ is reduced by decreasing $V_{th}$ and $\tau$. In addition, $\tau$ can be reduced by decreasing $C$ (see Supplementary Note A). $t_1$ has the largest effect among the three factors. $t_2$ and $t_3$ are related to the speed of light and the turn-on time of the MOSFET, respectively, and thus can be reduced by shortening the connection between the unit cells and using fast-response MOSFETs.

To demonstrate the performance of the proposed design, we simulated a simple proof-of-concept model using a hexagonal prism structure that included two interlinked MS unit cells and a receiver Rx, as shown in Fig. 3(a) (see Supplementary Note B for detailed design parameters). This structure included two adjacent conducting panels with symmetrical MS unit cells placed in front of two identical monopole transmitters Tx1 and Tx2 (18 mm long and spaced 120° apart) on a conducting ground surface. Tx1 and Tx2 generated a continuous wave and a pulsed sine wave at the same frequency of 3.10 GHz as the first and second (time-delayed multipath) signals, respectively. The Tx2 signal was generated with and without a 2-μs delay, which allowed MS1 to reach a steady state and send the bias to MS2. Both input powers were set to 30 dBm, which was sufficient to turn on the loaded diodes. A co-simulation method using an ANSYS Electronics Desktop Simulator (2022 R2) was adopted for the circuit design, simulation, and performance optimization (see Supplementary Note C).

When a signal was generated only by Tx1, a large voltage value was observed at Rx, as shown in Fig. 3(b) (blue curve). However, when both Tx1 and Tx2 were used to generate signals without any delay, the received voltage level was reduced by 11.5 dB due to the interference signal generated by Tx2. Note that the phase of the Tx2 signal was shifted by 180 degrees to easily distinguish the Tx1 and Tx2 signals (compare the orange and red curves to the blue curve in Fig. 3(b)). This interference was suppressed when the Tx2 signal, namely, the multipath signal, had a delay (purple curve). In this case, the $R_M$ value of MOSFET1 remained low due to the absence of bias from MS2. Therefore, a high potential was generated across $C_1$ in MS1, which was applied to bias the gate of MOSFET2, leading to the short circuiting of MSC2. For this reason, the transmittance of MS1 was higher than that of MS2; thus, in this case, the received voltage waveform was similar to the waveform obtained using only Tx1 (blue curve). The suppression performance is shown in Fig. 3(c), where each received voltage is presented on a polar plot. According to these results, the original Tx1 signal was reduced by only 2.3 dB and varied by -0.8 degrees, which indicates that the magnitude of the first incoming signal was enhanced by approximately 10 dB in our approach. Since the circuit structure is symmetrical, a similar response was obtained when the first wave (CW) was generated by Tx2 and the second wave (pulsed wave) was generated by Tx1. The signals shown in Fig. 3(b) contained third harmonics as well, although their magnitudes were at least 10 dB smaller than those of the fundamental mode (see Supplementary Note G for harmonic components).

A prototype for measurements was fabricated to test the feasibility of the proposed concept (see Supplementary Note D for the details of the fabricated prototype).

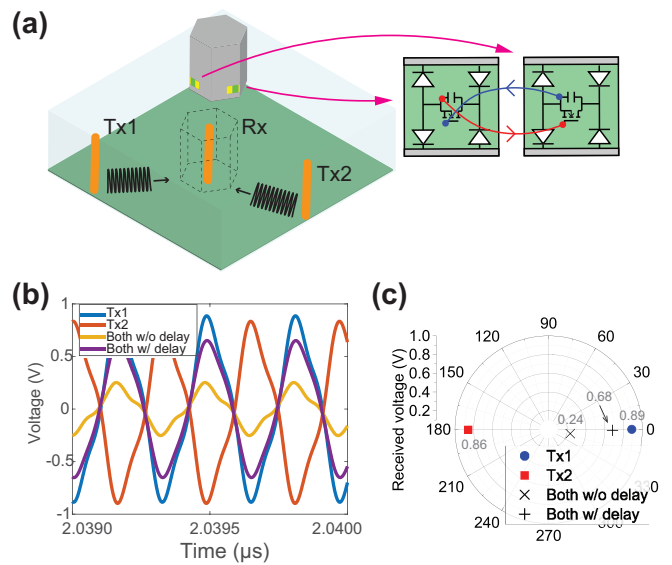

FIG. 3. Numerical validation. (a) Simulation model including Rx covered by an MS-based prism structure. The signals from Tx1 and Tx2 effectively mimicked the first signal and the multipath time-delayed signal, respectively. (b, c) The simulated received voltages in (b) the time domain and (c) the polar coordinate system. The polar plot shows the fundamental mode of the received voltages (see Supplementary Note G for harmonic components). The gray numbers near the symbols indicate the voltage values.

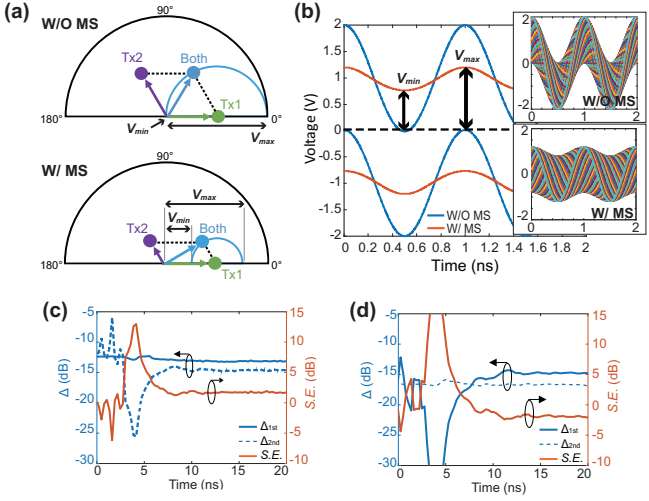

FIG. 4. Experimental validation. (a) The polar coordinate images of the received signals without (top panel) and with the MS (bottom panel). (b) The envelopes of the received signals, including the Tx1 (first) and Tx2 (second) signals in the time domain. (c, d) $\Delta$ and $S.E.$ when the first signal was generated by (c) Tx1 or (d) Tx2.

The phase and amplitude of the received voltage were evaluated, and the results are shown in Fig. 4. The measurements were conducted at 3.64 GHz, while they were performed at 3.10 GHz in the simulation; this change occurred because of the operation frequency shift resulting from fabrication inaccuracies and variations in the circuit parameters. First, we consider the case in which the first and second waves are generated by Tx1 and Tx2, respectively. Importantly, due to the challenge of fixing the phase difference between the two incident signals, the phase of the Tx1 signal was assumed to be $0°$, while that of the Tx2 signal was varied (i.e., unsynchronized). Fig. 4(a) shows polar phase images of the signal measured without and with the MS prism in the top and bottom panels, respectively. Since the signal received at Rx was determined by the superposition of the Tx1 and Tx2 signal vectors, the locus of the received signal was expected to have a reduced radius when the MS prism was used. This radius change was estimated from the envelopes of the time-domain signals received with and without the MS prism, as shown in Fig. 4(b); the insets in this figure show the received signal envelope without and with the MS prism (top and bottom insets, respectively). Based on these results, Fig. 4(b) presents the time-domain response of the received signal envelope, showing the maximum and minimum amplitudes, $V_{max}$ and $V_{min}$. These values were obtained without and with the MS prism (e.g., $V_{max}^{w/o}$ and $V_{max}^{w/}$). Although $V_{max}$ and $V_{min}$ were the only parameters available in our realistic measurements, $(V_{max}+V_{min})/2$ could be used to estimate the Tx1 signal magnitude according to Fig. 4(a) (note that the Tx1 vectors are determined by half of the $V_{max}$ and $V_{min}$ values in Fig. 4(a)). In addition, $(V_{max}-V_{min})/2$ represents the radius of the half-circle drawn by the superposition of the Tx1 and Tx2 signal vectors, which corresponds to the Tx2 signal magnitude. Therefore, the difference $\Delta_{1st}$ in the magnitudes of the first incoming waves without and with the MS prism was estimated as

$$\Delta_{1st} = \frac{V_{1st}^{w/}}{V_{1st}^{w/o}} = \frac{V_{max}^{w/} + V_{min}^{w/}}{V_{max}^{w/o} + V_{min}^{w/o}}, \quad (4)$$

where $V_{1st}$ denotes the magnitude of the voltage of the first incoming wave (i.e., $V_{1st}^{w/o}$ and $V_{1st}^{w/o}$ are the magnitudes of the first waves without and with the MS prism, respectively). Similarly, using $V_{2nd}$ as the magnitude of the second incoming wave without and with the MS prism ($V_{2nd}^{w/o}$ and $V_{2nd}^{w/}$, respectively), the difference $\Delta_{2nd}$ in the magnitudes of the second waves without and with the MS prism was evaluated by

$$\Delta_{2nd} = \frac{V_{2nd}^{w/}}{V_{2nd}^{w/o}} = \frac{V_{max}^{w/} - V_{min}^{w/}}{V_{max}^{w/o} - V_{min}^{w/o}}. \quad (5)$$

Thus, we evaluated the experimental shielding effectiveness $S.E.$ for multipath signals by dividing Eq. (4) by Eq. (5), i.e.,

$$S.E. = \Delta_{1st}/\Delta_{2nd}. \quad (6)$$

Note that Eqs. (4) to (6) are still valid even if the sources are exchanged between Tx1 and Tx2.

Fig. 4(c) shows plots of both $\Delta_{1st}$ and $\Delta_{2nd}$ as well as the corresponding $S.E.$ in the time domain when the first wave was generated by Tx1. Fig. 4(d) shows the same plots when the first wave was generated by Tx2. The results in Fig. 4(c) and Fig. 4(d) confirm that the second wave was suppressed regardless of the arrival direction of the radio waves. These results also indicate that the suppression performances were slightly different from each other, as the fabricated hexagonal prism was not perfectly symmetrical, as in the simulation model, due to the use of additional copper lines. Nonetheless, these results showed that the use of our prototype structure led to a suppression of more than 10 dB despite the use of signals with the same frequency and different signal orders, thus experimentally validating our proposed concept.

Next, we numerically extended the multipath wave suppression paradigm introduced in Fig. 3(a) by increasing the numbers of incident multipath signals and MS unit cells, as depicted in Fig. 5. First, a third transmitter Tx3 and a third MS unit cell MS3 were added. MS1, MS2, and MS3 unit cells were used for two of the six conducting panels in front of Tx1, Tx2, and Tx3, respectively, with each panel having $2 \times 3$ unit cells. As shown in Fig. 5(a), the three types of unit cells were interconnected (see also Supplementary Note F) and accepted only one of the three incident signals generated by Tx1, Tx2, and Tx3, which were separated by $120°$. The signals generated by Tx2 and Tx3 had a 2-$\mu$s delay so that MS1 could reach a steady state and bias the MOS-

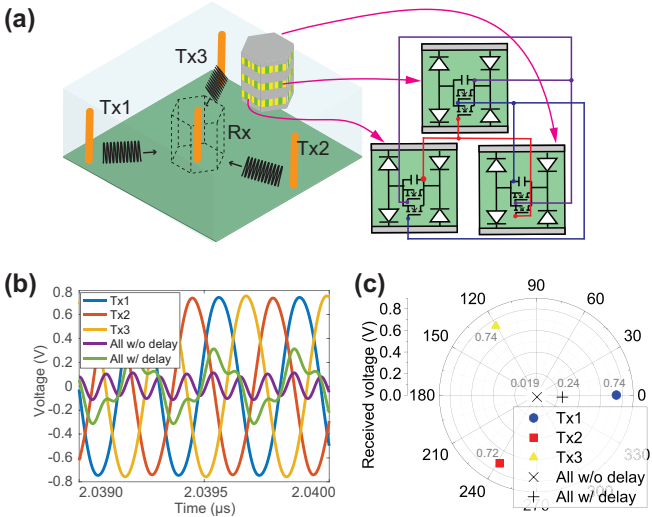

FIG. 5. Extended simulation model and results. (a) Entire simulation model. Sets of three interconnected MS unit cells were used. Each of the six conducting panels had $2 \times 3$ MS unit cells. The signal from Tx1 mimicked the first wave, while the signals from Tx2 and Tx3 mimicked multipath delayed signals. Each of the three signals passed through the interconnected MS unit cells. (b, c) The simulated received voltages in (b) the time domain and (c) the polar coordinate system. The polar plot shows the fundamental mode of the received voltages (see Supplementary Note G for harmonic components). The gray numbers near the symbols indicate the voltage values.

FETs in MS2 and MS3. The received signals are plotted in the time domain and the polar coordinate system in Fig. 5(b) and Fig. 5(c), respectively. These results show that the received voltage including the multipath time-delayed signals (from Tx2 and Tx3) was suppressed by approximately 9.7 dB with a phase change of -4.3 degrees. However, without the time delay (i.e., without the internal coupling mechanism), the magnitude of the received voltage was further reduced to 31.8 dB with a phase change of -69.3 degrees due to the 120° phase offset of the three signals. Thus, our results demonstrate that the magnitude of the first incoming signal was improved by more than 20 dB in our approach. Note that this case also showed third harmonics, which were approximately 10 dB smaller than the magnitude of the fundamental mode (see Supplementary Note G). In addition, we performed a simplified simulation Supplementary Note F, where only one set of three interconnected MS unit cells was used, which also validated our multipath suppression mechanism under a three-incident-wave scenario.

Our MS-based method effectively suppressed multipath signals while enabling the transmission of the first incoming wave despite the two physical limitations imposed by the LTI nature and the use of signals with the same frequency. In the literature, nonlinear MSs have been used as antennas [29], filters [30, 31], and reconfigurable intelligent surfaces (RISs) [32]. However, no prior studies have reported a passive approach for overcoming the two physical limitations imposed by the LTI nature, which limits the ability to transmit only the first incoming wave while eliminating delayed signals at the same frequency. Our method has advantages over other approaches based on modulation and signal processing techniques, as existing approaches can be implemented on advanced devices only but not on general low-cost devices. Nonetheless, the antenna design in our proof-of-concept prototype was simplified in this study; thus, its performance can be further improved. For instance, the first incoming wave was detected if the incident angle was distinguished, which depended on the number of MS unit cells employed. Thus, in a future work, the detectable incident angle range could be improved. Furthermore, the suppression performance was demonstrated to be approximately 10 dB but only for a limited time of approximately 1.6 ns. The suppression time may be extended by slowing the circuit mechanism that maintains an equilibrium state among the interconnected MOSFETs. Moreover, our proposed suppression effect was observed at three different times, i.e., $t_1$, $t_2$, and $t_3$. Thus, minimizing $t_1$, $t_2$, and $t_3$ is important for realizing fast-response multipath suppression in practice. Finally, although we focused on multipath issues in this study, our interlocking approach can be utilized to address other EM-related issues. For instance, antennas [29], sensors [33], imagers [34], RISs [35], and other microwave devices [36] can be autonomously controlled by an additional pulsed signal impinging from different directions.

In conclusion, we numerically and experimentally validated a filtering concept for spatial multipath waves to overcome the limitations imposed by the LTI nature of signals with the same frequency. Even with our passive approach, we can detect the direction of the first incoming wave while preventing the transmission of time-delayed interference signals with different angles but the same frequency. Our results show a suppression performance of more than 10 dB, which can be further improved by introducing an advanced antenna design. Our approach does not require DC supply voltages and is readily applicable in diverse situations, facilitating the implementation of complicated wireless communication setups. The idea of integrating same-frequency signal control strategies with spatially interlocked mechanisms can potentially be applied to develop advanced microwave devices and address existing EM issues.

This study was supported in part by the Japan Science and Technology Agency (JST) under the Precursory Research for Embryonic Science and Technology (PRESTO) and the Fusion Oriented Research for Disruptive Science and Technology and the National Institute of Information and Communications Technology (NICT), Japan under the commissioned research No. 06201.

[1] C. A. Balanis, *Antenna theory: analysis and design* (John Wiley & Sons, 2016).

[2] A. Goldsmith, *Wireless communications* (Cambridge university press, 2005).

[3] B. E. Henty and D. D. Stancil, *Multipath-enabled super-resolution for rf and microwave communication using phase-conjugate arrays*, Phy. Rev. Lett. **93**, 243904 (2004).

[4] J. Yeh, T. M. Antonsen, E. Ott, and S. M. Anlage, *First-principles model of time-dependent variations in transmission through a fluctuating scattering environment*, Phys. Rev. E **85**, 015202 (2012).

[5] Y. Yan, L. Li, G. Xie, C. Bao, P. Liao, H. Huang, Y. Ren, N. Ahmed, Z. Zhao, Z. Wang et al., *Multipath effects in millimetre-wave wireless communication using orbital angular momentum multiplexing*, Sci. Rep. **6**, 33482 (2016).

[6] H. Miyazawa and H. Itoh, *Development of a location system for tv ghost sources and its serviceability*, Electron. Commun. Jpn. **73**, 109 (1990).

[7] T. S. Rappaport, *Wireless communications: Principles and practice* (Pearson Education India, 2010).

[8] C. Balanis, *Advanced Engineering Electromagnetics* (John Wiley & Sons, 2012).

[9] N. Ida, *Engineering Electromagnetics* (Springer, 2004).

[10] W. Hayt and J. Buck, *Engineering Electromagnetics* (McGraw Hill, 2001).

[11] P. J. Antsaklis and A. N. Michel, *Linear systems* (Springer, 1997).

[12] J. Hong, *Microstrip filters for RF/Microwave Applications* (John Wiley & Sons, 2011).

[13] D. Sievenpiper, L. Zhang, R. Broas, N. Alexopolous, and E. Yablonovitch, *High-impedance electromagnetic surfaces with a forbidden frequency band*, IEEE Trans. Microw. Theory Techn. **47**, 2059 (1999).

[14] A. Li, S. Singh, and D. Sievenpiper, *Metasurfaces and their applications*, Nanophotonics, **7**, 989 (2018).

[15] N. Yu and F. Capasso, *Flat optics with designer metasurfaces*, Nat. Mater. **13**, 139 (2014).

[16] G. Y. Liu, L. Li, J. Q. Han, H. X. Liu, X. H. Gao, Y. Shi, and T. J. Cui, *Frequency-domain and spatial-domain reconfigurable metasurface*, ACS Appl. Mater. Interfaces **12**, 23554 (2020).

[17] S. Rout and S. Sonkusale, *Wireless multi-level terahertz amplitude modulator using active metamaterial-based spatial light modulation*, Opt. Express **24**, 14618 (2016).

[18] K. Wu, J. Liu, Y. Ding, W. Wang, B. Liang, and J. Cheng, *Metamaterial-based real-time communication with high information density by multipath twisting of acoustic wave*, Nat. Commun. **13**, 5171 (2022).

[19] B. Assouar, B. Liang, Y. Wu, Y. Li, J. Cheng, and Y. Jing, *Acoustic metasurfaces*, Nat. Rev. Mater. **3**, 460 (2018).

[20] G. Ma and P. Sheng, *Acoustic metamaterials: From local resonances to broad horizons*, Sci. Adv. **2**, e1501595 (2016).

[21] C. Pfeiffer and A. Grbic, *Metamaterial huygens' surfaces: tailoring wave fronts with reflectionless sheets*, Phy. Rev. Lett. **110**, 197401 (2013).

[22] N. Yu, P. Genevet, M. A. Kats, F. Aieta, J.-P. Tetienne, F. Capasso, and Z. Gaburro, *Light propagation with phase discontinuities: generalized laws of reflection and refraction*, Science **334**, 333 (2011).

[23] M. Lapine, I. V. Shadrivov, and Y. S. Kivshar, *Colloquium: nonlinear metamaterials*, Rev. Mod. Phys. **86**, 1093 (2014).

[24] A. D Boardman, V. V. Grimalsky, Y. S. Kivshar, S. V Koshevaya, M. Lapine, N. M. Litchinitser, V. N. Malnev, M. Noginov, Y. G. Rapoport, and V. M. Shalaev, *Active and tunable metamaterials*, Laser Photonics Rev. **5**, 287 (2011).

[25] H. Wakatsuchi, D. Anzai, J. J. Rushton, F. Gao, S. Kim, and D. Sievenpiper, *Waveform selectivity at the same frequency*, Sci. Rep. **5**, 9639 (2015).

[26] H. Takeshita, D. Nita, Y. Cheng, A. A. Fathnan, and H. Wakatsuchi, *Dual-band waveform-selective metasurfaces for reflection suppression*, Appl. Phys. Lett. **123**, 191703 (2023).

[27] H. Wakatsuchi, J. Long, and D. F. Sievenpiper, *Waveform selective surfaces*, Adv. Funct. Mater. **29**, 1806386 (2019).

[28] K. Asano, T. Nakasha, and H. Wakatsuchi, *Simplified equivalent circuit approach for designing time-domain responses of waveform-selective metasurfaces*, Appl. Phys. Lett. **116**, 171603 (2020).

[29] D. Ushikoshi, R. Higashiura, K Tachi, A. Aminulloh Fathnan, S. Mahmood, H. Takeshita, H. Homma, R. Akram, S. Vellucci, J. Lee et al., *Pulse-driven self-reconfigurable meta-antennas*, Nat. Comm. **14**, 633 (2023).

[30] H. Wakatsuchi, *Time-domain filtering of metasurfaces*, Sci. Rep. **5**, 16737 (2015).

[31] K. Takimoto, H. Takeshita, A. A. Fathnan, D. Anzai, S. Sugiura, and H. Wakatsuchi, *Perfect pulse filtering under simultaneous incidence at the same frequencies with waveform-selective metasurfaces*, APL Mater. **11**, 81116 (2023).

[32] M. Sharma, M. Tal, C. McDonnell, and T. Ellenbogen, *Electrically and all-optically switchable nonlocal nonlinear metasurfaces*, Sci. Adv. **9**, eadh2353 (2023).

[33] M. Beruete and I. Jáuregui-López, *Terahertz sensing based on metasurfaces*, Adv. Optical Mater. **8**, 1900721 (2020).

[34] L. Li, Y. Shuang, Q. Ma, H. Li, H. Zhao, M. Wei, C. Liu, C. Hao, C. Qiu, and T. Cui, *Intelligent metasurface imager and recognizer*, Light Sci. Appl. **8**, 97 (2019).

[35] E. Ayanoglu, F. Capolino, and A. Swindlehurst, *Wave-controlled metasurface-based reconfigurable intelligent surfaces*, IEEE Wirel. Commun. **29**, 86 (2022).

[36] M. Barbuto, D. Lione, A. Monti, S. Vellucci, F. Bilotti, and A. Toscano, *Waveguide components and aperture antennas with frequency-and time-domain selectivity properties*, IEEE Trans. Antennas Propag. **68**, 7196 (2020).

*Supplementary Material for*

**Multipath Signal-Selective Metasurface: Passive Time-Varying Interlocking Mechanism to Vary Spatial Impedance for Signals with the Same Frequency**

Kaito Tachi, Kota Suzuki, Kairi Takimoto, Shunsuke Saruwatari, Kiichi Niitsu, Peter Njogu, and Hiroki Wakatsuchi

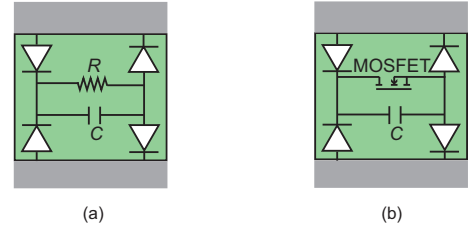

FIG. S1. MS unit cell design. (a, b) MS unit cells using a pair consisting of a capacitor and (a) a resistor or (b) a MOSFET.

*List of Supplementary Notes*

**Note A:** The fundamental design and mechanism of MS unit cells

**Note B:** The design of the hexagonal prism structure and the antenna

**Note C:** Co-simulation method

**Note D:** Fabrication of the hexagonal prism structure

**Note E:** The measurement setup and method

**Note F:** The hexagonal prism structure with three interconnected unit cells

**Note G:** Harmonic components in the simulated hexagonal prism structure

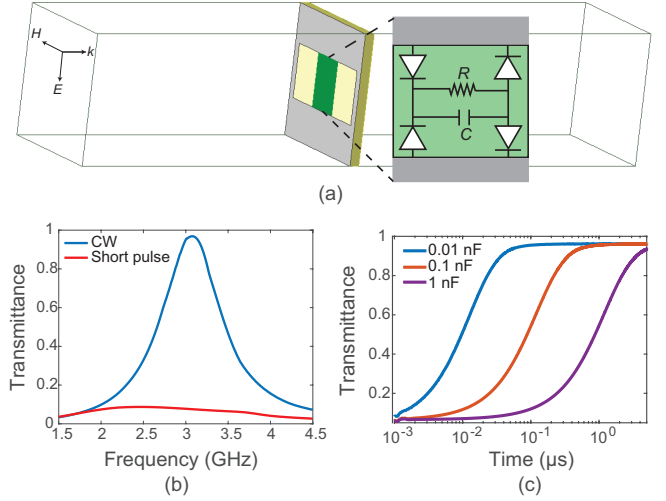

FIG. S2. Transmittance of the MS using the unit cell shown in Fig. S1a. (a) The simulation model. The periodicity was 17 mm, and the slit dimensions were set to 15 mm and 7 mm for the incident electric and magnetic field directions, respectively (i.e., the same as the design parameters adopted in Fig. 2(c)). (b) The frequency-domain profile for an input power of 10 dB. (c) The time-domain profiles at 3.1 GHz with various $C$ values. $R$ was fixed at 1 MΩ. The simulation method is explained in Supplementary Note C.

### Supplementary Note A: The fundamental design and mechanism of MS unit cells

In this supplementary note, we explain the design and mechanism of the unit cells on which the MS used in this work was based. First, we designed a unit cell composed of a capacitor $C$ and a resistor $R$ connected in parallel within a diode bridge, as shown in Fig. S1(a) [1]. Inside the diode bridge, the frequency of the EM wave impinging on the MS was converted to an infinite set of components due to the rectification effect of the diode bridge, although most of the incident energy was converted to zero frequency. Therefore, similar to the transients of classic direct current (DC) circuits, this unit cell setup allowed control of the rectified charges through the time-domain response of the paired $C$ and $R$. Specifically, the electric potential of the capacitor was gradually increased, while the number of incoming electric charges entering the diode bridge was decreased. Therefore, the transmittance of the short pulse signal was limited. In contrast, a CW signal was effectively transmitted, as the capacitor was fully charged and the intrinsic resonant mechanism of the MS slit was maintained. This structure is numerically demonstrated in Fig. S2(a) using periodic boundaries for the directions of the incident electric and magnetic fields (see Supplementary Note C for the simulation method). Fig. S2(b) shows that the transmittance of CWs was higher than that of short pulses, even for signals at the same frequency of approximately 3.0 GHz, which can be explained by the above-mentioned mechanism. In addition, the results in Fig. S2(c) demonstrate how the capacitance value influenced the time-domain response; the higher the capacitance was, the larger the time constant became, as observed in classic transient circuits. This figure indicates that by reducing the capacitance value, the transmitting state of the unit cell can be quickly changed, which is suitable for reducing the response time for multipath suppression in this study.

Importantly, to realize multipath suppression, an additional mechanism was needed to transmit only the first incoming wave while not allowing the transmission of any delayed signals with the same frequency. This was achieved by replacing the resistor $R$ with a MOSFET, as shown in Fig. S1(b). First, in this capacitor $C$-MOSFET configuration, the MOSFET acted as a variable resistor, as its effective resistive component $R_M$ between the drain and source varied with its gate-source voltage $V_g$. Specifically, when $V_g$ became larger than the threshold voltage of the MOSFET, the state between the drain and source switched from an open circuit to a short circuit, namely, $R_M$ decreased. Thus, when $R_M$ was high, an incoming signal was permitted to pass through the MS since the intrinsic resonant mechanism of the MS slit was maintained, as explained by the transient response of the unit cell with the paired $RC$ circuit. However, when $R_M$ was low, the transient response disappeared, as $C$ could no longer effectively store the rectified electric charges, leading to poor transmittance of the incoming wave. In the antenna configuration demonstrated in Fig. 3(a), $V_g$ was increased by the addition of the neighboring MS unit cell that received the first incoming wave. However, $V_g$ remained low until a signal arrived at the MS unit cells. Additionally, in a practical multipath suppression mechanism, $C$ should be reduced to minimize the response time, as explained with $t_1$ in the main manuscript. Compared to the unit cell design in Fig. 2(c), $C_{add}$ was not included in the design shown in Fig. S1(b). However, note that the use of $C_{add}$ contributes to characterizing only the frequency-domain response and not the time-domain response (see Eq. (2) and [2]).

### Supplementary Note B: The design of the hexagonal prism structure and the antenna

This supplementary note provides the design parameters necessary for the MS-based hexagonal prism structure shown in Fig. 3(a). Table S1 presents the dimensions of the proposed hexagonal prism shown in Fig. S3. Fig. S3(a) shows the MS slit of the structure in Fig. 2(c), while Fig. S3(b) and Fig. S3(c) represent the hexagonal prism (without any slit) and the internal monopole antenna, respectively, which are related to Fig. 3(a) and Fig. 5(a). The SPICE parameters for the MOSFET are shown in Table S2, while those of the bridge diodes are shown in Table S3. Additionally, Table S4 presents the values of $C$ and $C_{add}$.

### Supplementary Note C: Co-simulation method

This supplementary note describes the co-simulation method linking the EM model with the lumped components in the circuit simulator. ANSYS Electronics Desktop 2022 R2 was used to model and simulate the structures shown in Figs. 2, 3, and 5. First, an EM model without any lumped circuit elements was simulated; these elements were replaced with lumped ports in HFSS, the EM solver of ANSYS. The calculation results were then imported into the circuit simulator as

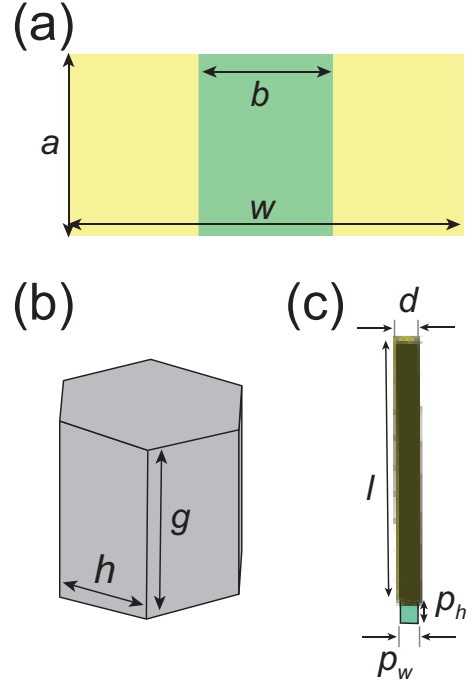

FIG. S3. The design of the various components in the proposed spatial MS filters shown in Fig. 3. (a) The MS slit, (b) the hexagonal structure, and (c) the monopole. The design dimensions are given in Table S1. The substrate of the MS and the hexagonal prism is a 1.27 mm thick Rogers 3010 material.

TABLE S1. Design parameters of the proposed MS-based hexagonal prism shown in Fig. S3.

| Parameter | Length (mm) |
|---|---|
| $a$ | 7 |
| $b$ | 5 |
| $w$ | 15 |
| $h$ | 35 |
| $g$ | 51 |
| $l$ | 18 |
| $d$ | 1 |
| $p_h$ | 0.5 |
| $p_w$ | 1 |

TABLE S2. SPICE parameters used for MOSFETs (Toshiba, 2SK1062).

| Characteristics | | Parameter | Value | Units |
|---|---|---|---|---|
| Drain-source voltage | | $V_{DS}$ | 60 | V |
| Gate-source voltage | | $V_{GSS}$ | +/-20 | V |
| Drain current | DC | $I_D$ | 200 | mA |
| | Pulse | $I_{DP}$ | 800 | |
| Drain power dissipation($Ta$=25°C) | | $P_D$ | 200 | mW |
| Channel temperature | | $T_{ch}$ | 150 | °C |
| Storage temperature range | | $T_{stg}$ | -55 to 150 | °C |

a circuit model. Importantly, by connecting the circuit model with actual circuit elements via lumped ports, our MS and MS-based prism were simulated with markedly

TABLE S3. SPICE parameters for diodes (Avago, HSMS-286x series).

| Parameter | Units | Value |
|---|---|---|
| $B_V$ | V | 7.0 |
| $C_{J0}$ | pF | 0.18 |
| $E_G$ | eV | 0.69 |
| $I_{BV}$ | A | 1e-5 |
| $I_S$ | A | 5e-8 |
| $N$ | | 1.08 |
| $R_S$ | $\Omega$ | 6.0 |
| $P_B$ (VJ) | V | 0.65 |
| $P_T$(XTI) | | 2 |
| $M$ | | 0.5 |

TABLE S4. Capacitance values adopted in Fig. 2(b).

| Parameter | Value |
|---|---|
| $C$ | 100 pF |
| $C_{add}$ | 2 pF |

reduced simulation times, which helped us find suitable design parameters.

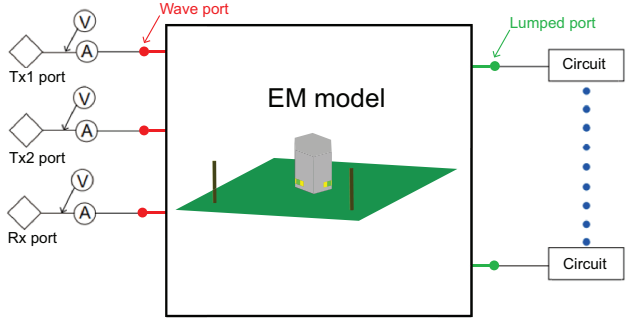

FIG. S4. Circuit schematic image for the co-simulation method.

### Supplementary Note D: Fabrication of the hexagonal prism structure

This supplementary note explains the method for fabricating the MS-based hexagonal prism. Fig. S5(a) shows the ground plane, the two transmitting monopole antennas, and the receiving monopole antenna, while Fig. S5(b) shows the receiving monopole covered by the MS-based hexagonal prism. The hexagonal structure consisted of six identical conducting panels with Rogers 3010 substrates. Their copper cladding surfaces (or the conducting surfaces of the MS unit cells) were deployed to face external directions. The six panels were tightly fixed by copper tape so that the signal passed through only the MS unit cells (i.e., slits). A hexagonal metallic plate was used to cover the top of the six assembled panels, and the entire structure (i.e., the entire hexagonal prism) was deployed on a copper ground plane, with the receiving monopole antenna set at the center position of the prism. All monopole antennas were grounded and connected to coaxial cables via holes, as shown in Fig. S5(a). These design parameters were set to be the same as the parameters adopted in the simulations (see Supplementary Note B).

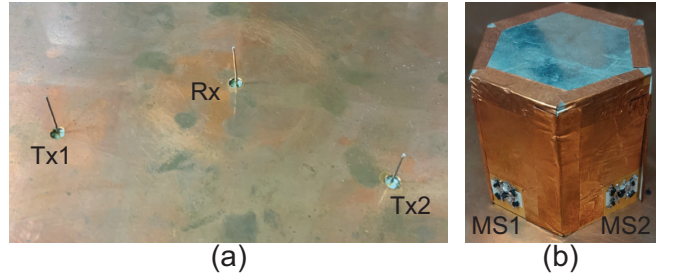

FIG. S5. The fabricated MS-based hexagonal prism. (a) The ground plane, the two transmitting monopole antennas, and the receiving monopole antenna. (b) The receiver covered by the MS-based hexagonal prism antenna on the ground plane.

### Supplementary Note E: The measurement setup and method

This supplementary note explains how the measurement setup was configured to obtain the measurement results shown in Fig. 4. Our setup consisted of two signal generators (Anritsu, MG369C and Keysight Technologies, N5183B) to generate first and second (delayed) waves as a CW and a 2-$\mu$s pulsed wave, respectively. Both signals had the same frequency of 3.64 GHz and the same input power of 30 dBm, which was obtained by applying amplifiers (Ophir, 5193RF and Mini Circuits, HPA-50W-63+). Initially, the first signal was generated by Tx1, while the second signal was generated by Tx2, as shown in Fig. S6(a). Next, the two signal sources were exchanged so that the first and second waves were generated by Tx2 and Tx1, respectively, as shown in Fig. S6(b). In both cases, part of the second signal energy was sent to an oscilloscope (Keysight Technologies, UXR0134A) via a coupler (ET Industry, C-058-30) to observe how the receiver accepted the two signals when the second signal arrived. In contrast to the simulation results in Fig. 3, the phase difference between the two signals could not be easily fixed in the measurements. Therefore, the phase difference varied in our measurements. Thus, we alternatively measured the varying amplitude of the envelope made by the two unsynchronized signals, which could be applied to evaluate the multipath suppression performance, as explained in Fig. 4(a) and Eqs. (4) to (6).

Fig. S6(c) and Fig. S6(d) show the amplitude of the received signal and the amplitude envelope for received signals without and with the MS hexagonal prism struc-ture, respectively, when the first wave was generated by Tx1. Fig. S6(e) and Fig. S6(f) show the same results but with the signal sources exchanged (i.e., the first and second waves were generated by Tx2 and Tx1, respectively). The resultant magnitude differences $\Delta_{1st}$ and $\Delta_{2nd}$ between the first and second signals in the first setup (Fig. S6(a)) were obtained from Fig. S6(c) and Fig. S6(d). The magnitude differences $\Delta_{1st}$ and $\Delta_{2nd}$ for the second setup (Fig. S6(b)) were obtained from Fig. S6(e) and Fig. S6(f). The plots of $\Delta_{1st}$ and $\Delta_{2nd}$ and their differences (i.e., $S.E.$) are shown in Fig. 4(c) and Fig. 4(d) for the first and second setups, respectively.

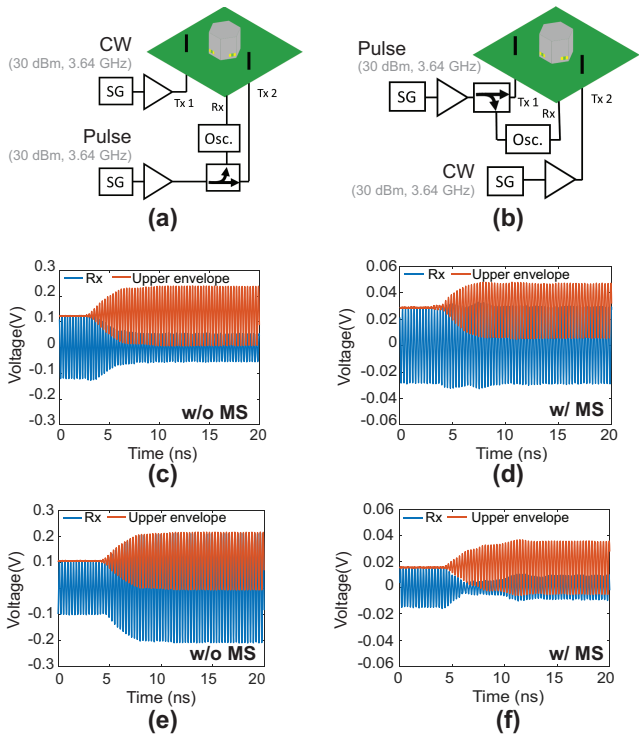

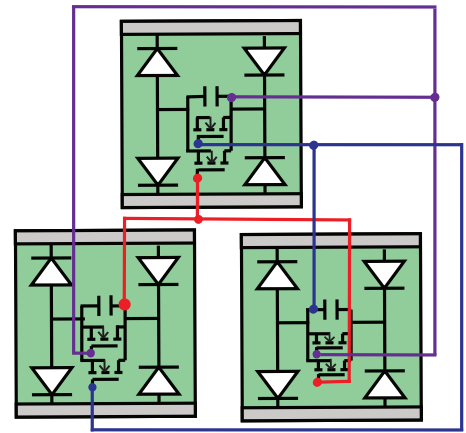

FIG. S7. The interconnections among the three MS unit cells shown in Fig. 5(a).

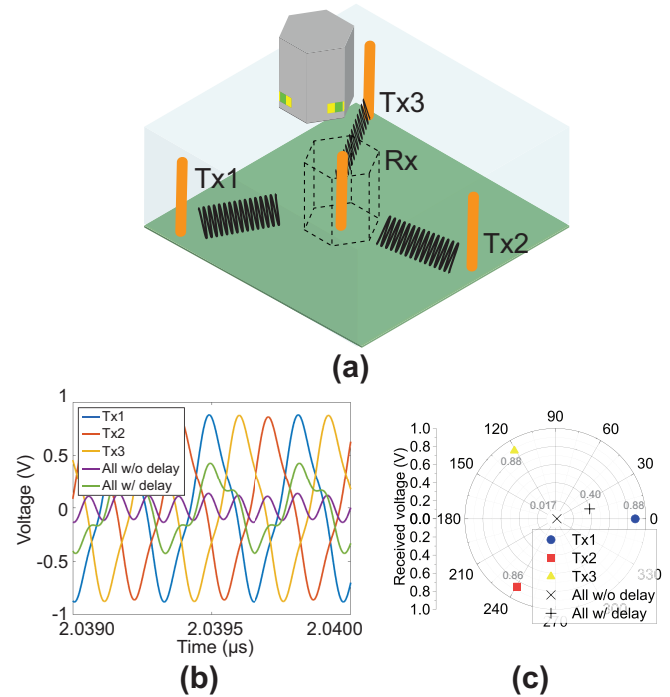

FIG. S6. The measurement setup for the MS-based hexagonal prism structure. (a) The first setup. The first and second signals were generated by Tx1 and Tx2, respectively. (b) The second setup. The first and second signals were generated by Tx2 and Tx1, respectively. (c, d) The received signals and upper envelopes for the first setup (c) without and (d) with the MS-based hexagonal prism. (e, f) The received signals and upper envelopes for the second setup (e) without and (f) with the MS-based hexagonal prism.

### Supplementary Note F: The hexagonal prism structure with three interconnected unit cells

This supplementary note explains the interconnections among the MS unit cells shown in Fig. 5. Each cell consisted of a diode bridge, a capacitor, and two MOSFETs, as shown in Fig. S7. The second MOSFET was used here because three incident sources (i.e., three incident paths) were considered to generate the results shown in Fig. 5.

FIG. S8. Simplified simulation results using only one set of three interconnected unit cells. (a) Schematic image of the simulation. (b, c) The simulated received signals in (b) the time domain and (c) the polar coordinate system. The polar plot shows the fundamental mode of the received voltages (see Supplementary Note G for harmonic components). The gray numbers near the symbols indicate the voltage values.

The capacitor and the two MOSFETs were connected in parallel within the diode bridge. To interlink three types of cells responding to the three different waves, the capacitor in each cell was connected to the gate of the MOSFETs in the other two cells, as depicted in Fig. S7.

With this approach, the potential of the capacitor biased the two MOSFETs, changing the impedance of the two unit cells and preventing the transmission of the two delayed signals.

Using the interconnections shown in Fig. S7, where each of the hexagonal panels had $2 \times 3$ unit cells, we performed simulations, and the simulation results are presented in Fig. 5. Simplified simulations were also performed, as shown in Fig. S8. In this case, the total number of unit cells was limited to only three (namely, only one set of three interconnected cells), as shown in Fig. S8(a). Compared to the results in Fig. 5, the multipath suppression performance was improved, as shown in Fig. S8(b) and Fig. S8(c), as the three unit cells were spaced more in this setup, which reduced the influence of other incoming waves.

**Supplementary Note G: Harmonic components in the simulated hexagonal prism structure**

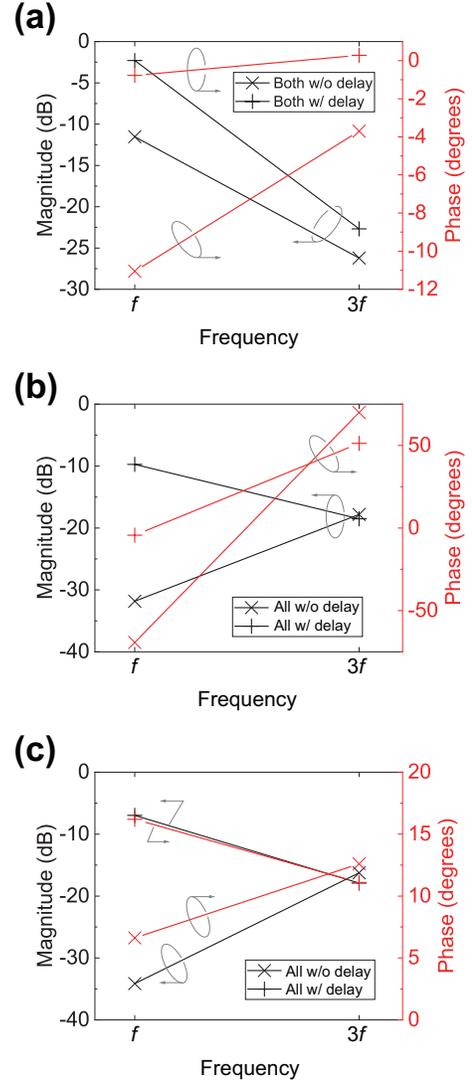

FIG. S9. The fundamental mode and the third harmonic components in the simulated hexagonal prism structures in Fig. 3, Fig. 5, and Fig. S8. (a-c) The fourier-transformed results of (a) Fig. 3(b), (b) Fig. 5(b), and (c) Fig. S8(b). $f$ represents the fundamental mode, while $3f$ is the third harmonic. $f = 3.1$ GHz. The black and the red symbols correspond to the left and right axes, respectively. The magnitudes are normalized to the Tx1 signal in each simulation, while the phase changes show the difference from the phase of the Tx1 signal.

In the simulation models shown in Fig. 3, Fig. 5, and Fig. S8, the fundamental mode of interference signals was successfully suppressed in Fig. 3(c), Fig. 5(c), and Fig. S8(c), although high harmonic components appeared in the time-domain voltages of Fig. 3(b), Fig. 5(b), and Fig. S8(b), which is clarified in this supplementary note. In Figs. S9(a), S9(b), and S9(c), we show both the fundamental mode and the third harmonic components fourier-transformed from Fig. 3(b), Fig. 5(b), and Fig. S8(b), respectively. According to Fig. S9(a), the magnitudes of the third harmonics were at least 10 dB smaller than those of the fundamental modes (see the black symbols). In Fig. S9(b) and Fig. S9(c), the magnitudes of the third harmonics were higher than the magnitude of the fundamental mode if there was no delay in the interference signals. With the time delay, however, the fundamental mode showed largest magnitudes among the others. Note that these harmonic signals can be readily eliminated by introducing a low-pass filter. Therefore, our approach can effectively suppress the interference signals while transmitting the first incoming signals.

## REFERENCES

[1] H. Wakatsuchi, J. Long, and D. F. Sievenpiper, *Waveform selective surfaces*, Adv. Funct. Mater. **29**, 1806386 (2019).

[2] K. Asano, T. Nakasha, and H. Wakatsuchi. *Simplified equivalent circuit approach for designing time-domain responses of waveform-selective metasurfaces*, Appl. Phys. Lett. **116**, 171603 (2020).